\documentclass[10pt,a4paper,twocolumn,superscriptaddress,aps,nofootinbib]{revtex4}
\usepackage{amsmath,amssymb,graphicx}
\makeatletter
\usepackage[active]{srcltx}
\usepackage{graphicx,color}
\usepackage{changebar}
\usepackage{hyperref}
\hypersetup{
	colorlinks=true,
	urlcolor= blue,
	citecolor=blue,
	linkcolor= blue,
	bookmarks=true,
	bookmarksopen=false,
}
\usepackage[T1]{fontenc}
\usepackage{ae}
\usepackage[ansinew]{inputenc}
\usepackage{amsmath}
\usepackage{braket}
\usepackage{mathtools}
\usepackage{slashed}
\usepackage{empheq}
\usepackage{tikz}
\usepackage{multirow}
\usetikzlibrary{decorations.pathmorphing}
\usepackage{amsmath,amssymb,graphics,epsfig,subfigure}
 \usetikzlibrary{quotes,angles}
 \usepackage{etoolbox}
 
\usepackage{graphicx}
\usepackage{amsfonts}
\usepackage{amsmath}
\usetikzlibrary{arrows} 
\usetikzlibrary{decorations.markings}

\newcommand{\orcidicon}{%
	\begin{tikzpicture}
	\draw[lime, fill=lime] (0,0) 
	circle [radius=0.15] 
	node[white] {{\fontfamily{qag}\selectfont \tiny ID}};
	\draw[white, fill=white] (-0.0625,0.095) 
	circle [radius=0.007];
	\end{tikzpicture}	\hspace{-2mm}
}
\newcommand\orcidAdriano{{\href{https://orcid.org/0000-0003-1871-2068}{\orcidicon}}}
\newcommand\orcidCarlos{{\href{https://orcid.org/0000-0001-6913-0223}{\orcidicon}}}
\newcommand\orcidRicardo{{\href{https://orcid.org/0000-0001-8802-3634}{\orcidicon}}}

\begin{document}

\title{ Holonomy corrected Schwarzschild black hole lensing}

\author{A. R. Soares\orcidAdriano\!}
\email{adriano.soares@ifma.edu.br }
\affiliation{Grupo de Estudos e Pesquisas em Laborat\'orio de Educaç\~ao matem\'atica, Instituto Federal de Educa\c{c}\~ao Ci\^encia e Tecnologia do Maranh\~ao,  R. Dep. Gast\~ao Vieira, 1000, CEP 65393-000 Buriticupu, MA, Brazil.}

\author{C. F. S. Pereira \orcidCarlos\!}
\email{carlos.f.pereira@edu.ufes.br}
\affiliation{Departamento de F\'isica e Qu\'imica, Universidade Federal do Esp\'irito Santo, Av.Fernando Ferrari, 514, Goiabeiras, Vit\'oria, ES 29060-900, Brazil}

\author{R. L. L. Vit\'oria \orcidRicardo\!}
\email{ricardo.vitoria@pq.cnpq.br/ricardo-luis91@hotmail.com}
\affiliation{Faculdade de F\'isica, Universidade Federal do Par\'a, Av. Augusto Corr\^ea, Guam\'a, 66075-110, Bel\'em, PA, Brazil.}

\author{Erick Melo Rocha}
\email{ erick.rocha@ifma.edu.br}
\affiliation{Instituto Federal de Educaç\~ao Ci\^encia e Tecnologia do Maranh\~ao, Campus Buriticupu, CEP 65393-000, Buriticupu, Maranh\~ao, Brazil.}
\affiliation{MUST University Florida, 2220 N Federal Hwy, Boca Raton, FL 33431, USA. }

\begin{abstract}
	
In the present work, we theoretically investigate gravitational lensing in the spacetime of a holonomy corrected Schwarzschild black hole. Analytical expressions for the light deflection angle are obtained in both the weak field limit and the strong field limit. Furthermore, we analyze observables, such as relativistic images and magnifications, and compare the results with those expected in a Schwarzschild spacetime. We discuss the possibilities and difficulties of investigating such a solution in practice. 
	
\end{abstract}



\maketitle

\section{Introduction}
It is well known that the limitations presented by Einstein's General Relativity in dealing naturally with the problem of geodesic singularity have motivated the search for a broader gravitation theory \cite{1457,529,tst, mod-gra-rep,fr theo, felice, lb, report, mod}.
From this perspective, Loop Quantum Gravity (LQG) emerges as one of the promising quantum gravity proposals with the objective of ``regularizing" gravity \cite{Rovelli003, Ashtekar84, Perez80}. LQG is a non-pertubative theory to quantize the spacetime structure; however, it does not present a complete quantum description close to what we currently understand as a singularity. In this sense, one can explore the effects of LQG in low-energy regimes through effective models, that is, models with corrections arising from quantum effects. Recently, in \cite{Alonso-Bardaji829, Alonso-Bardaji106}, the authors obtained a solution of a regular, geodesically complete black hole, resulting from anomaly-free holonomy corrections. Some aspects of this model have already been investigated , such as those linked to quasi-normal modes \cite{Zhang08421, Moreira107} and horizon area \cite{Sobrinho14}. In this work, specifically, we theoretically investigate aspects linked to gravitational lensing related to the aforementioned spacetime in order to provide theoretical tools with which can observationally investigate the plausibility of the theory.

Gravitational lensing is a phenomenon resulting from the interaction between light and gravity. When light passes through the gravitational field of a given source, it changes its direction of propagation as a result of this interaction \cite{Einstein1936, Liebes}. If the light passes through a region very far from the source that generated the gravitational field,  the
angular deflection  is small and we call this regime the weak field limit. Conversely, as the distance between the source and the light decreases,  the
angular deflection increases. In fact, if the light touches a certain minimum distance, the deflection angle diverges, we call this regime the strong field limit. The first studies of lensing in the strong field limit, in the 1960s, showed that relativistic images are too weak \cite{Darwin249,Atkinson517}, in addition, there were no lensing equations suitable for this regime. Because of this, the field of research remained at a standstill in the following decades. In \cite{Virbhadra-Ellis-2000}, Virbhadra and Ellis presented lensing equations suitable for studying lensing in the strong field regime. Later, Bozza developed a suitable methodology to derive the angular deflection of light in the strong field limit \cite{Bozza2002}, which was later improved by Tsukamoto \cite{Tsukamoto2017}.
Since then, gravitational lensing has been investigated in several contexts involving black holes \cite{Eiroa2002, Eiroa2004, Tsukamoto201734, rotacao, Aazami, azquez, Bozza-Sereno, Bozza-Scarpetta, VBz,Virbhadra-064038, Virbhadra-2204.01792}, wormholes \cite{Nakajima85, Chetouani, Nandi, Dey-sen, Gibbons-Vyska, Tsukamoto-Harada, Nandi-Potapov, Tsukamoto2017-95, Tsukamoto-Harada-2017, Shaikh-Banerjee, Gao2211.17065, Huang2302.13704,Abe2010}, topological defects \cite{Cheng, Sharif2015, Cheng92, Cheng28}, modified theories of gravitation \cite{Sotani92, Wei75, Bhadra, Eiroa73, Sarkar23, Mukherjee39, Gyulchev75, Chen80, Shaikh96} and regular black holes \cite{Eiroa28,Eiroa88,soaresBbounce, Ghosh006}.

Combined with these theoretical developments, recent technological advances, such as the Event Horizon Telescope (EHT) \cite{AkiyamaL1,AkiyamaL2,AkiyamaL3,AkiyamaL4,AkiyamaL5,AkiyamaL6}, have strongly activated the field of lensing research in the strong field limit. The next generation Event Horizon Telescope (ngEHT) \cite{EHT2} is expected to be able to discriminate between relativistic images and the first and second images. This, among other achievements, presents itself as an important tool for distinguishing between observations predicted by General relativity and other theories of gravity, such as LQG, within the high energy regime.

Motivated by this stimulating context, in this work, we theoretically explore the observational signature in the gravitational lensing of a solution arising from LQG. We compare the results with those predicted by Scharzschild spacetime in order to distinguish between the two cases and discuss the results. The work is organized as follows. In Section \ref{sec2} we present the metric that describes the spacetime and calculate the light deflection  in the weak field limit. In section \ref{sec3}, we analytically obtain an expression for the light deflection in the strong field limit. In section \ref{sec4} we derive the observables in the strong field limit and compare them with those in the Schwarzschild spacetime. We also discuss the possibilities of experimental verification of the solution. In the \ref{conc} section, we concluded.

\section{Spacetime and Light deflection}\label{sec2}
\begin{figure}[h]
	\centering
	\includegraphics[width=\columnwidth]{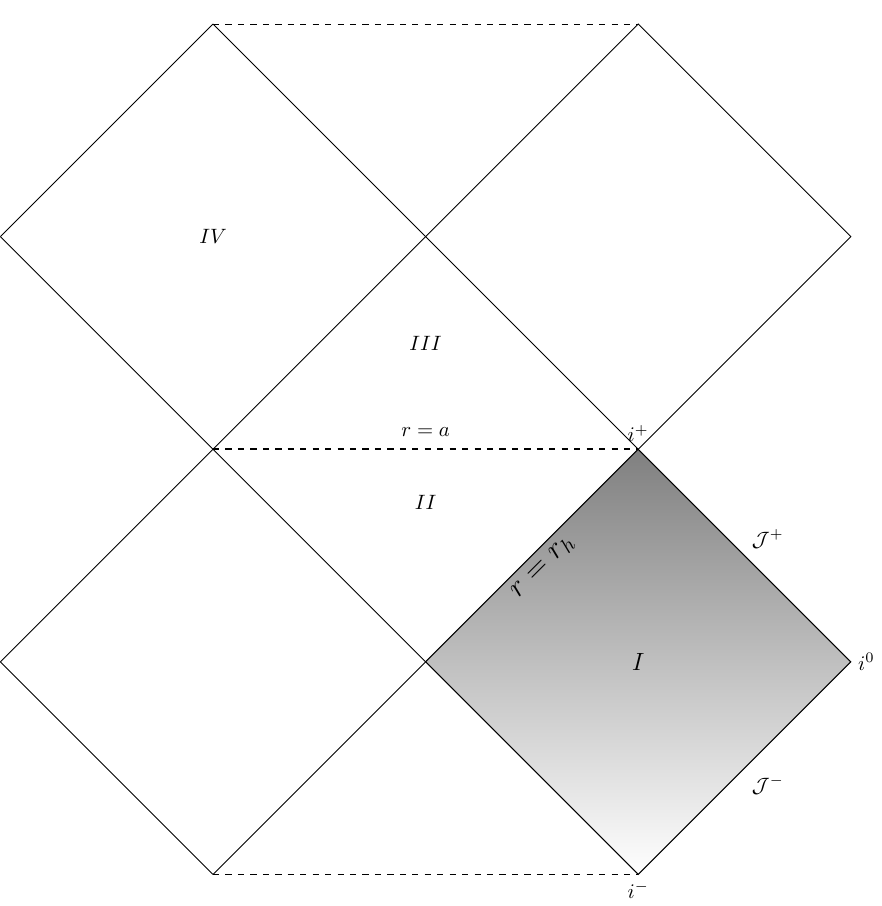}
	\caption{ Penrose diagram. We are analyzing lensing in the shaded region $I$, beyond to the hypersurface $r=r_{h}$. The other regions correspond to the black hole (II), white hole (III) and another asymptotically flat region (IV) \cite{Alonso-Bardaji106}.  } 
	\label{pen}
\end{figure}

The line element metric that describes the holonomy corrected Schwarzschild black hole, in spherical coordinates, is given by \cite{Alonso-Bardaji829, Alonso-Bardaji106}
\begin{eqnarray}\label{me}
ds^2&=&-\bigg(1-\frac{2M}{r}\bigg)dt^2+\frac{r}{r-a}\bigg(1-\frac{2M}{r}\bigg) ^{-1}dr^2\nonumber\\
&& + r^2\left(d\theta^2+\sin^2\theta d\phi^2\right) \ .
\end{eqnarray}
In (\ref{me}), the LQG parameter, $a$, is such that $a<2M$. It is a static, spherically symmetric and asymptotically flat spacetime. The spacetime given in (\ref{me}) has a wormhole-like structure, with $a$ defining a minimal spacelike hypersurface separating the trapped regular BH
interior from the anti-trapped other region. However, we will consider light rays that do not cross the event horizon, characterized by $r=r_{h}=2M$, i.e., we will address regions where $r> r_{h}$, considering the global structure in the form of a Penrose diagram (see Fig.\ref{pen}), we will work within region I .

Having established the limits of our investigation, let us derive the equations necessary to study lensing. Using the variational method, the Lagrangian corresponding to the minimum distance in the spacetime described in (\ref{me}) , for $\theta=\frac{\pi}{2}$, is given by
\begin{eqnarray}\label{l}
	\mathcal{L}&=&-\bigg(1-\frac{2M}{r}\bigg)\bigg(	\frac{dt}{d\lambda}\bigg)^2+\frac{r}{r-a}\bigg(1-\frac{2M}{r}\bigg)^{-1}\bigg(\frac{dr}{d\lambda}\bigg)^2\nonumber\\
	&& +r^2\bigg(\frac{d\phi}{d\lambda}\bigg)^2 \ .
\end{eqnarray}
The corresponding Euler-Lagrange equation for the coordinates $t$ and $\phi$ leads to the following conserved quantities
\begin{equation}\label{en}
E=\bigg(1-\frac{2M}{r}\bigg)\bigg(\frac{dt}{d\lambda}\bigg) \ ,
\end{equation}
and
\begin{equation}\label{mom}
L=r^2\frac{d\phi}{d\lambda} \ .
\end{equation}
which can be understood as energy and angular momentum. Replacing (\ref{en}) and (\ref{mom}) into (\ref{l}) and considering null geodesics, where $\mathcal{L}=0 $, the (\ref{l}) leads to

\begin{equation}\label{eq1}
\frac{r}{r-a}\bigg(\frac{dr}{d\lambda}\bigg)^2=E^2-\frac{L^2}{r^2}\bigg(1-\frac {2M}{r}\bigg) \ .
\end{equation}
Eq.(\ref{eq1}) can be seen as describing the dynamics of a classical particle of energy $E$ subject to an effective potential $V_{eff}=\frac{L^2}{r^2}\bigg(1-\frac{2M}{r}\bigg)$. The value of $r$ for which the orbits are circular is determined by $\frac{dV_{eff}(r)}{dr}=0$, with this we conclude that the radius photon sphere, $r_m$, in the holonomy corrected Schwarzschild BH is given by
\begin{equation}\label{eq2}
r_m= 3M \ ,
\end{equation}
just like in Schwarschild spacetime.

To analyze the spacetime lensing of a holonomy corrected Schwarzschild BH, let us consider a photon departing from the asymptotically flat region and approaching the BH at a radial distance $r_{0}$, called the turning point, such that $r_{ 0}>r_m$. After being deflected by the BH's gravitational field, the photon heads to another asymptotically flat region. At the turning point, $E=V_{eff}(r_{0})$, that is,

\begin{equation}\label{eq3}
	\frac{1}{\beta^2}=\frac{1}{r_{0}^2}\bigg(1-\frac{2M}{r_{0}}\bigg) \ ,
\end{equation}
where $\beta(r_{0})=\frac{L}{E}$ is the impact parameter. Replacing (\ref{mom}) in (\ref{eq1}), we find
\begin{equation}\label{eq4}
\Big(\frac{d\phi}{dr}\Big)=\bigg[\bigg(1-\frac{a}{r}\bigg)\bigg[\frac{r^4}{\beta^ 2}-r^2\bigg(1-\frac{2M}{r}\bigg)\bigg]\bigg]^{-1/2} \ .
\end{equation}
We want to find the change in coordinate $\phi$, i.e., $\Delta\phi= \phi_{-}-\phi_{+}$. By symmetry, the contributions to $\Delta\phi$ before and after the turning point are equal, so
Eq.(\ref{eq4}) leads to
\begin{eqnarray}\label{eq55}
\Delta\phi&=&2\int_{r_{0}}^{\infty}\bigg[\bigg(1-\frac{a}{r}\bigg)\bigg[\frac{r^4}{\beta^2}\bigg.\nonumber\\
&& \big.-r^2\bigg(1-\frac{2M}{r}\bigg)\bigg]\bigg]^{-1/2} dr \ .
\end{eqnarray}
Let's introduce the following variable change
$u=\frac{1}{r}$, where we have $dr=-\frac{du}{u^2}$. Furthermore, $u\to0$ when $r\to\infty$ and $u\to u_0$ when $r\to r_0$. Therefore, in terms of $u$, (\ref{eq55}) becomes
\begin{equation}\label{eq5}
\Delta\phi=2\int_{0}^{u_{0}}\bigg[(1-au)\bigg[\frac{1}{\beta^2}-u^2(1-2Mu)\bigg]\bigg]^{-1/2} dr\ .
\end{equation}
From (\ref{eq3}), to (\ref{eq5}) becomes
\begin{equation}\label{eq6}
\Delta\phi=2\int_{0}^{u_{0}}\bigg[(1-au)\bigg[ u_0^2(1-2Mu_0)-u^2(1-2Mu)\bigg]\bigg]^{-1/2} dr\ .
\end{equation}
In the weak field approximation, that is, assuming that the photon passes very far from the BH, we can take the approximation $M\ll1$ and $a\ll1$. Therefore, up to the second order in $a$, a (\ref{eq6}) provides the deflection of light $\alpha=\Delta\phi-\pi$:
\begin{equation}\label{eq7}
\alpha\simeq\frac{4M}{\beta}+\frac{a}{\beta} +\frac{3\pi a^2}{16\beta^2} +\frac{Ma(3\pi -4)}{4\beta^2} \ .
\end{equation}
In Eq.(\ref{eq7}), the first term refers to the deflection of a Schwarszchild BH, while the following terms bring a contribution from the Holonomic correction. Later, we will study the observational implications of these corrections.
  
  \section{Expansion for deflection of light in the strong field limit}\label{sec3}
  
 In this section, we will obtain the expression for the deflection of light in the strong field limit. To do this, we will adopt the methodology developed in \cite{Bozza2002,Tsukamoto2017}. Making the following variable change
 \begin{equation}\label{eq8}
 z=1-\frac{r_0}{r} \ ,
 \end{equation}
 the (\ref{eq55}) becomes
 \begin{equation}\label{eq9}
 I(r_0)=\int_{0}^{1} \frac{2r_0}{\sqrt{G(z,r_0)}}\ dz \ ,
 \end{equation}
 where,
 \begin{eqnarray}\label{eq10}
 G(z,r_0)&=&\frac{r_0^4}{\beta^2}-\frac{ar_0^3}{\beta^2}(1-z)-r_0^2(1-z) ^2 \nonumber\\
 &&+(2M+a)r_0(1-z)^3\nonumber\\
 &&-2Ma(1-z)^4 \ .
 \end{eqnarray}
 Expanding $G(z,r_0)$ in power series, we obtain
 \begin{equation}\label{eq11}
 G(z,r_0)\simeq\alpha_1(r_0)z+\alpha_2(r_0)z^2 \ ,
 \end{equation}
 with
 \begin{equation}\label{eq12}
 \alpha_1(r_0)=6aM-2ar_0-6Mr_0+2r_0^2
 \end{equation}
 and
 \begin{equation}\label{eq13}
 \alpha_2(r_0)=3ar_0+6Mr_0-12aM-r_0^2 \ .
 \end{equation}
 In the strong field limit, $r_0\to r_m=3M$, the expansion coefficients are
 \begin{equation}\label{eq14}
 \alpha_1(r_0)\to\alpha_1(r_m)=0
 \end{equation}
 and
 \begin{equation}
 \alpha_2(r_0)\to\alpha_2(r_m)=9M^2-3aM \ .
 \end{equation}
 This demonstrates that the integral (\ref{eq9}), in dominant order, diverges logarithmically in the strong field limit. In order to obtain an expression for the deflection of light in the strong field limit, we will divide (\ref{eq9}) into two parts, a divergent part $I_D(r_0)$ and a regular part $I_R(r_0)$ , where
 \begin{equation}\label{eq15}
 I_R(r_0)=I(r_0)-I_D(r_0) \ .
 \end{equation}
 The divergent part is given by
 \begin{equation}
 I_D(r_0)=\int_{0}^{1} \frac{2r_0}{\sqrt{\alpha_1(r_0)z+\alpha_2(r_0)z^2}}\ dz \ ,
 \end{equation}
 whose integration provides
 \begin{eqnarray}\label{eq16}
 I_D(r_0)&=&-\frac{4r_0}{\sqrt{\alpha_2(r_0)}}\log(\sqrt{\alpha_1(r_0)})\nonumber\\
 && +\frac{4r_0}{\sqrt{\alpha_2(r_0)}}\log\bigg(\sqrt{\alpha_2(r_0)}\bigg.\nonumber\\
 &&\bigg.+\sqrt{\alpha_1(r_0)+\alpha_2(r_0)}\bigg) \ .
 \end{eqnarray}
 Now let us expand $\alpha_1(r_0)$ and $\beta(r_0)$ close to the radius of the photon sphere. From (\ref{eq12}) and (\ref{eq3}), we are left with
 \begin{equation}\label{eq17}
 \alpha_1(r_0)\simeq(6M-2a)(r_0-3M) \
 \end{equation}
 and
 \begin{equation}\label{eq18}
 \beta(r_0)\simeq\sqrt{27M^2}+\sqrt{\frac{3}{4M^2}}(r_0-3M)^2 \ .
 \end{equation}
 From (\ref{eq17}) and (\ref{eq18}), we have that
 \begin{equation}\label{eq19}
 \alpha_1(r_0)\simeq2\sqrt{6}(3M^2-aM)\sqrt{\bigg(\frac{\beta}{\sqrt{27M^2}}-1\bigg)} \ .
 \end{equation}
 Replacing (\ref{eq19}) in (\ref{eq16}) and considering the strong field limit, that is, $r_0\to r_m=3M$, we get
 \begin{eqnarray}\label{eq20}
 I_D&=&-\sqrt{\frac{3M}{3M-a}}\log\bigg(\frac{\beta}{3\sqrt{3}M}-1\bigg)\nonumber\\
 &&+\sqrt{\frac{3M}{3M-a}}\log(6) \ .
 \end{eqnarray}
 The regular part, (\ref{eq15}), is given by
 \begin{eqnarray}\label{eq21}
 I(r_0)&=&\int_{0}^{1} \frac{2r_0}{\sqrt{G(z,r_0)}}\ dz \nonumber\\
 &&-\int_{0}^{1} \frac{2r_0}{\sqrt{\alpha_1(r_0)z+\alpha_2(r_0)z^2}}\ dz \ .
 \end{eqnarray}
 In the limit $r_0\to r_m=3M$ and taking into account (\ref{eq10}), (\ref{eq11}) and (\ref{eq3}), (\ref{eq21}) leads to
 \begin{eqnarray} \label{eq22}
 I_R&=&\sqrt{\frac{12}{3-2(a/2M)}} \times\nonumber\\
 &&\log\bigg(\frac{18-12(a/2M)}{6-(a/2M)+3\sqrt{3-2(a/2M)}}\bigg) \ .
 \end{eqnarray}
 Therefore, putting together (\ref{eq22}) and (\ref{eq20}), we finally find the expansion for the deflection of light in the strong field limit,
 \begin{eqnarray}\label{eq23}
 \alpha(\beta)&=&-\sqrt{\frac{3M}{3M-a}}\log\bigg(\frac{\beta}{3\sqrt{3}M}-1\bigg) \nonumber\\
 &&+\sqrt{\frac{3M}{3M-a}}\log(6) +\sqrt{\frac{12}{3-2(a/2M)}}\times\nonumber\\
 &&\log\bigg(\frac{18-12(a/2M)}{6-(a/2M)+3\sqrt{3-2(a/2M)}}\bigg)\nonumber\\
 && -\pi \ .
 \end{eqnarray}
 One can easily check that when the LQG parameter, $a$, tends to zero, the deflection falls into the well-known expression referring to the Schwarszchild spacetime \cite{Bozza2002}:
  \begin{equation}
  \alpha(\beta)=-\log\left(\frac{b}{3\sqrt{3}M}-1\right)+\log(6)+0.9496-\pi \ .
  \end{equation}
   
\begin{figure}[h]
  	\centering
  	\includegraphics[width=\columnwidth]{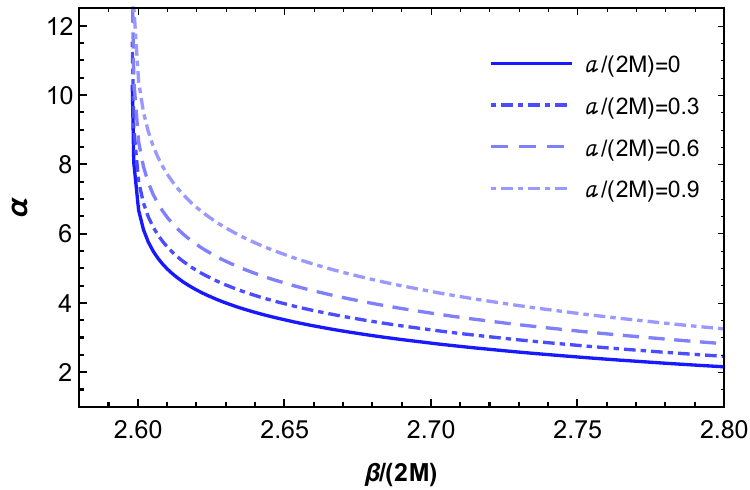}
  	\caption{Light angular deflection  as a function of the impact
  		parameter $\beta/2M$ for various values of the LQG parameter $a/2M$.} 
  	\label{LG}
  \end{figure}

\begin{figure}[h]
   	\centering
   	\includegraphics[width=\columnwidth]{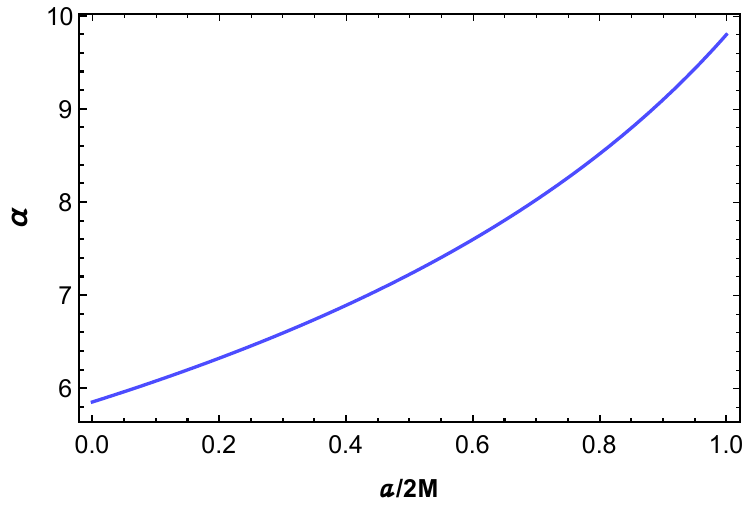}
   	\caption{Light angular deflection as a function of  $a/2M$ for $\frac{\beta}{2M}=\frac{3\sqrt{3}}{2}+0.005$ \ .} 
   	\label{LG2}
   \end{figure}

 In Fig.\ref{LG} we plot the light deflection for some values of $a/2M$, we see that it is a function that increases with LQG parameter. The solid curve corresponds to the well-known Schwarzschild spacetime case.
  In Fig.\ref{LG2}, we plot $\alpha$ as a function of $a/2M$ for $\frac{\beta}{2M}=\frac{3\sqrt{3}}{2}+0.005$ , that is, we are considering the deflection for a small radial displacement from the critical impact parameter as a function of the LQG parameter. As we can see more clearly, the deflection is a function that increases with the LQG parameter $a/2M$. This already indicates that the higher the LQG parameter, the more expressive the discrimination between the deviation of light in comparison to the Schwarzschild spacetime, which really demonstrates the coherence of the results.
  
   \section{Observables}\label{sec4}
   
   Now that we have obtained the expressions for the deflection of light in the weak field limit (\ref{eq7}) and in the strong field limit (\ref{eq23}), let's investigate, through gravitational lensing, what to expect observationally from the LQG spacetime. First, we will briefly review the lens equations in the strong field limit, then we will use them to derive expressions for the observables, with which we can observationally investigate the plausibility of the solution and distinguish it from the Schwarzschild black hole.
   
\begin{figure}[h]
    	\centering
    	\includegraphics[width=\columnwidth]{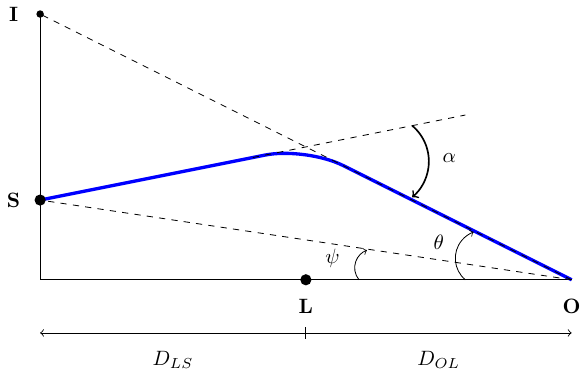}
    	\caption{Light angular deflection  diagram.} 
    	\label{fig3}
    \end{figure}

    The visual profile of the lensing is schematized in Fig. (\ref{fig3}). The light that is emitted by the source $S$ is deflected towards the observed $O$ by the {\bf LQG} compact object located in $L$. The angular deflection of light is given by $\alpha$. The angular positions of the source and image in relation to the optical axis, $\overline{LO}$, are given, respectively, by $\psi$ and $\theta$.
    
    As in \cite{Boz-Cap2001,Virbhadra-Ellis-2000}, we will admit that the source ($S$) is almost perfectly aligned with the lens ($L$) which is where relativistic images are most expressive. In this case, the lens equation relating the angular positions $\theta$ and $\beta$ is given by
    \begin{equation}\label{EqLente}
    \psi=\theta-\frac{D_{LS}}{D_{OS}}\Delta\alpha_{n}\ ,
    \end{equation}
    where $\Delta\alpha_n$ is the least deflection angle of all the loops made by the photons before reaching the observer, that is, $\Delta\alpha_{n}=\alpha-2n\pi$. In this approximation,
    \begin{equation}\label{betaap}
    \beta\simeq\theta D_{OL} \ ,
    \end{equation}
    so that, for simplicity, we can write the angular deflection (\ref{eq23}) as
    \begin{equation}\label{defle}
    \alpha(\theta)=-\bar{a}\log\left(\frac{\theta D_{OL}}{\beta_c}-1\right)+\bar{b}\ ,
    \end{equation}
    where, compared to (\ref{eq23}), we have
    \begin{equation}
    \bar{a}=\sqrt{\frac{3M}{3M-a}} \ ,
    \end{equation}
    \begin{eqnarray}
    \bar{b}&=&\sqrt{\frac{3M}{3M-a}}\log(6) +\sqrt{\frac{12}{3-2(a/2M)}}\times\nonumber\\
    &&\log\bigg(\frac{18-12(a/2M)}{6-(a/2M)+3\sqrt{3-2(a/2M)}}\bigg)\nonumber\\
    && -\pi \ ,
    \end{eqnarray}
    and
    \begin{equation}
    \beta_{c}=3\sqrt{3}M \ .
    \end{equation}
    
    What enters the lens equation is $\Delta\alpha_{n}$, to obtain it we expand $\alpha(\theta)$ close to $\theta=\theta^{0}_n$, where $\alpha (\theta^{0}_n)=2n\pi$. Thus, we are left with
    \begin{equation}\label{da}
    \Delta\alpha_n=\frac{\partial\alpha}{\partial\theta}\Bigg|_{\theta=\theta^0_n}(\theta-\theta^0_n) \ .
    \end{equation}
    Evaluating (\ref{defle}) in $\theta=\theta^{0}_n$, we obtain
    \begin{equation}\label{To}
    \theta^0_{n}=\frac{\beta_c}{D_{OL}}\left(1+e_n\right), \qquad\text{where}\quad e_n=e^{\frac{\bar{ b}-2n\pi}{\bar{a}}} \ .
    \end{equation}
    Substituting (\ref{defle}) and (\ref{To}) into (\ref{da}), we get
    \begin{equation}\label{de}
    \Delta\alpha_n=-\frac{\bar{a}D_{OL}}{\beta_ce_n}(\theta-\theta^0_n) \ .
    \end{equation}
    Substituting (\ref{de}) in the lens equation (\ref{EqLente}), we obtain the expression for the nth angular position of the image
    \begin{equation}\label{tetan}
    \theta_n\simeq\theta^0_n+\frac{\beta_ce_n}{\bar{a}}\frac{D_{OS}}{D_{OL}D_{LS}}(\psi-\theta^0_n) \ .
    \end{equation}
   Although the deflection of light preserves surface brightness, gravitational lensing changes the appearance of the source's solid angle. The total flux received by a lensed image is proportional to the magnification $\mu_{n}$, which is given by $ \mu_n=\left|\frac{\psi}{\theta}\frac{\partial\psi}{ \partial\theta}|_{\theta^0_{n}}\right|^{-1}$. Then from (\ref{EqLente}) and (\ref{de}) , we get
    \begin{equation}
    \mu_{n}=\frac{e_n(1+e_n)}{\bar{a}\psi}\frac{D_{OS}}{D_{LS}}\left(\frac{\beta_c}{D_{OL}}\right)^2 \ .
    \end{equation}
   In fact, $\mu_n$ decreases very quickly with $n$, so the brightness of the first image $\theta_1$ dominates in comparison with other ones. On the other hand, whatever the case, the presence of the factor $\left(\frac{\beta_c}{D_{OL}}\right)^2$ implies that the magnification will always be small. It should also be noted that at the limit of the source alignment, the lens and observer the magnification diverges, maximizing the possibility of detecting relativistic images.
   \subsection{Observables in the strong field limit}
   Finally, we have expressed the position of the relativistic images as well as their flows in terms of the expansion coefficients ($\bar{a}$, $\bar{b}$, and $\beta_c$). Let us now consider the inverse problem, that is, from observations, determine the expansion coefficients. With this, we can understand the nature of the object that generates the gravitational lens and compare it with the predictions made by modified theories of gravity.
   
\begin{figure}
     	\centering
     	\includegraphics[width=\columnwidth]{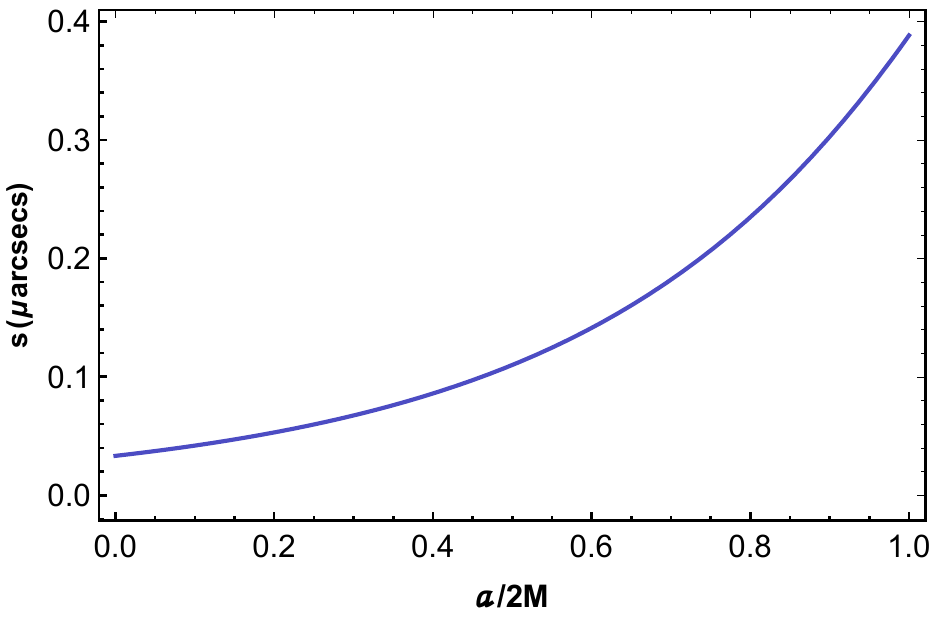}
     	\caption{ Angular separation $s$ as a function of $a/2M$} 
     	\label{fig4}
     \end{figure}

\begin{figure}
      	\centering
      	\includegraphics[width=\columnwidth]{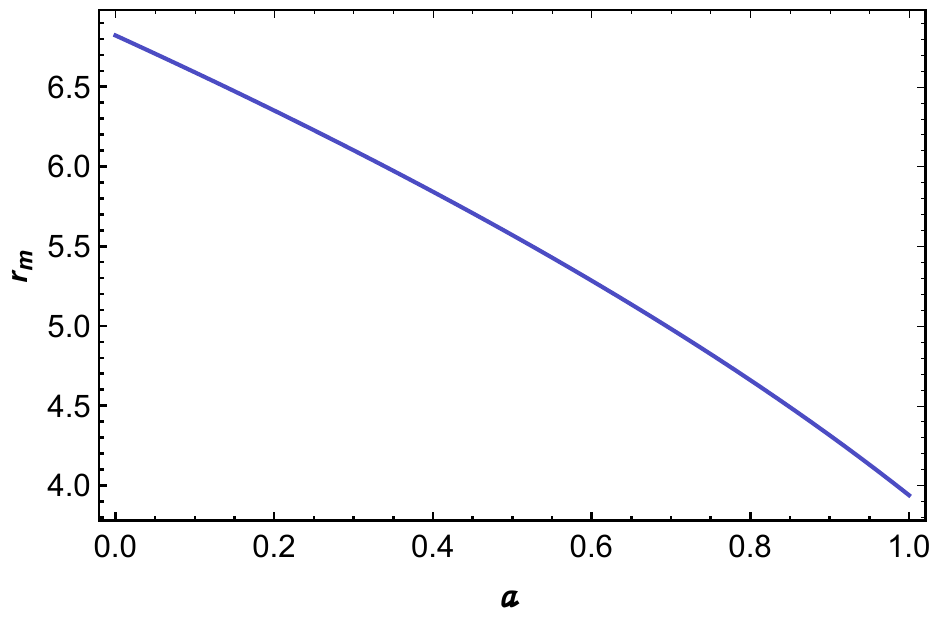}
      	\caption{$2.5\log_{10}\tilde{r}$ as a function of $a/2M$} 
      	\label{fig5}
      \end{figure}

   From the equations (\ref{tetan}) and (\ref{To}), we have that
    \begin{equation}
    	\theta_{\infty}=\theta_{\infty}^{0}=\frac{\beta_c}{D_{OL}} \ .
    \end{equation}
Therefore, we can express the critical impact parameter as
   \begin{equation}
   \beta_c=D_{OL}\theta_{\infty} \ .
   \end{equation}
   Let's follow Bozza \cite{Bozza2002} and assume that only the outermost image $\theta_{1}$ is discriminated as a single image while the rest are encapsulated in $\theta_{\infty}$. Therefore, Bozza defined the following observables,
   \begin{eqnarray}
   s&=&\theta_{1}-\theta_{\infty}= \theta_{\infty} e^{\frac{\bar{b}-2\pi}{\bar{a}}},\\
   \tilde{r}&=& \frac{\mu_{1}}{\sum_{n=2}^{\infty} \mu_{n} }= e^{\frac{2\pi}{\bar {a}}} \ .
   \end{eqnarray}
   In the expressions above, $s$ is the angular separation and $\tilde{r}$ is the relationship between the flow of the first image and the flow of all others. These forms can be inverted to obtain the expansion coefficients. To evaluate the observables, let us consider that the object in question has an estimated mass of $4.4\times10^{6}M_{\odot}$ and is at an approximate distance of $D_{OL}=8.5$Kpc, these data are the same for the black hole at the center of our galaxy \cite{Gezel2010}. As $\beta_c=3\sqrt{3}M$ does not depend on $a$, we can calculate it directly. In geometric units, where $M\to M\frac{G}{c^2}$, we will have $\theta_{\infty}\simeq26,5473\mu$arcsecs; which is the same as the Schwarzschild case, therefore, it is more promising to investigate the other parameters. We plot the behavior of the angular separation in Fig.\ref{fig4} as a function of $\frac{a}{2M}$. In Fig.\ref{fig5}, we plot
    \begin{equation}
    	r_m=2.5\log_{10}\tilde{r}
    \end{equation} 
   in function of $\frac{a}{2M}$.
  As we can see, the angular separation increases with $a$, while $r_m$ decreases. In Table \ref{t0}, we present the observables for some values of the parameter $\frac{a}{2M}$, where $\frac{a}{2M}=0$ corresponds to the Schwarzschild case. We observe that for $\frac{a}{2M}\ge0.6$ the angular separation increases by an order of magnitude compared to the Schwarzschild case. Given the increase in optical resolution of observational projects \cite{EHT2}, we hope in the coming decades to have minimum conditions to discriminate different theories of gravitation, however, it is still a great challenge. As we can still see from the graphs and tables, the ratio between the flow of the first and other images decreases, indicating an increase in the magnitude of the other images. 
  
      \begin{table}[]
      	\caption{observables}
      	\label{t0}
      	\begin{tabular}{|l|l|l|}
      		\hline
      		$a/(2M)$ & s($\mu$ arcsecs)          & $r_{m}$(magnitudes) \\ \hline
      		$0$        & $0.0332$    & $6.8218$              \\ \hline
      		$0.3$   & $0.0673$  &  $6.1017$               \\ \hline
      		$0.6$    & $0.1412$   & $5.2842$                \\ \hline
      		$0.9$ & $0.3026$   & $4.3145$  \\ \hline
      	\end{tabular}
      \end{table}
   \subsection{Observables in the weak field limit}
  The weak field limit corresponds to a very large impact parameter ($\beta \gg M$), so that there are no loops performed by light. In this limit, Eq. (\ref{eq7}) gives
   \begin{equation}\label{B1}
   	\alpha\sim\frac{4M+a}{\beta} \ .
   \end{equation}
   In the perfect alignment between the source, the compact object and the observer, we have $\psi=0$, this implies, from Eq. (\ref{EqLente}), that
   \begin{equation}\label{B2}
    \theta=\frac{D_{LS}}{D_{OS}}\Delta\alpha_{n}\ ,
   \end{equation}
  with $\Delta\alpha_{n}$ given by (\ref{B1}). Therefore, substituting (\ref{B1}) into (\ref{B2}), we find 
   \begin{equation}\label{B3}
   	\theta=\theta_{E}=\sqrt{\frac{D_{LS}(4M+a)}{D_{OS}D_{OL}}} \ ,
   \end{equation}
  where $\theta_{E}$ is the angular position of the Einstein ring. Beforehand, it is clear that the presence of the LQG parameter, $a$, increases the angular position of the Einstein ring.
  From (\ref{betaap}), we can also calculate the radius of the Einstein ring, $R_{E}$, which is given by
   \begin{equation}
   R_{E}=D_{OL}\theta_{E} \ .
   \end{equation}
    Let us now estimate the observables $R_E$ and  $\theta_E$ taking into account reasonable values for the LQG parâmetro. As in \cite{Abe2010}, let's consider the lensing of  a bulge star.  The following parameters are adopted: $D_{OS}=8$Kpc and $D_{OL}=4$Kpc. Based on these estimates, in the Table \ref{t1},  we present some values of the observables for the  presented scenario. We verified that, within the possibilities of the LQG gravity, the present solution presents feasible theoretical possibilities of being detected since the measurements are within the range of observables.
   \begin{table}[]
   	\caption{Einstein Radii/angle for Bulge  Lensing}
   	\label{t1}
   	\begin{tabular}{|l|l|l|}
   		\hline
   		$\frac{a}{2M}$ & $R_E$ (km)             & $\theta_E$(arcsecs) \\ \hline
   		$0$        & $1.26\times10^{12}$    &      $2.11$       \\ \hline
   		$0.3 $   & $1.35\times 10^{12}$  &  $2.26$               \\ \hline
   		$0.6$    & $1.44\times 10^{12}$   & $2.41$                \\ \hline
   		$0.9$ & $1.52\times 10^{12}$   & $ 2.54 $  \\ \hline
   	\end{tabular}
   \end{table}\\
 Let us make some observations about the limitation of this work. To calculate the observables, we use the conventional formulation for the lens equation (Eq. \ref{EqLente}). A general formulation for a spherically symmetric and static spacetime is given in \cite{Bozza103005} and improved by Takizawa \textit{et al.} \cite{Takizawa064060}, in order to solve the lens equation through iterative methods. In \cite{Takizawa064060}, the method is described clearly and in detail, but in order to analyze second-order corrections in holonomy correction, let us give a brief summary of the method here. In the iterative scheme, angles are considered small and expanded in terms of an iterative parameter ($\varepsilon$): $\psi=\varepsilon\psi_{(1)}$, $\theta=\sum_{k=1} ^{\infty}\varepsilon^k\theta_{(k)}$, $\alpha=\sum_{k=1}^{\infty}\varepsilon^k\alpha_{(k)}$\ . Substituting these expressions into the general lens equation gives the lens equation in all orders. For example, in the leading order ($\mathcal{O}(\varepsilon)$), you get the conventional lens equation:
  \begin{equation}
  	\psi_{(1)}=\theta_{(1)}-\frac{D_{LS}}{D_{OS}}\alpha_{(1)} \ .
  \end{equation}
 	At second order ($\mathcal{O}(\varepsilon^2)$), the general lens equation gives
 	\begin{equation}\label{t2}
 		\theta_{(2)}=\frac{D_{LS}}{D_{OS}}\alpha_{(2)} \ ,
 	\end{equation} 
 and so on. Next, we outline the steps that allow us to calculate $\theta_{(2)}$. To do so, let us write $\beta=\varepsilon D_{OL}+\varepsilon^2 D_{OL}+\mathcal{O}(\varepsilon^3)$, the mass $M\to\varepsilon^2M $ and $a\to\varepsilon^2a$. Substituting these values into (\ref{eq7}) we obtain
  \begin{equation}\label{2o}
  	\alpha_{(2)}=\frac{3a\pi-16D_{OL}\theta_{(2)}}{16D_{OL}^2\theta_{(1)}}(a+4M)-\frac{aM}{D_{OL}^2\theta_{(1)}^2} \ .
  \end{equation}
 Replacing (\ref{2o}) in (\ref{t2}), we find
  \begin{equation}\label{theta2}
  	\theta_{(2)}=\frac{3a\pi}{32D_{OL}}-\frac{aM}{2D_{OL}(a+4M)} \ .
  \end{equation}
  
With the expression (\ref{theta2}), we estimate the values of $\theta_{(2)}$ for $0 <a/2M<1$; in all cases we find that $\theta_{(2)}\sim 10^{-6}$arcsecs. This contribution is far below the observational range, which justifies, in the present case, considering the conventional lens equation (leading order in $\varepsilon$).
 
\section{Conclusions}\label{conc}
In this work, we theoretically investigate the gravitational lensing produced by a recently proposed spacetime that is motivated by LQG. We provide analytical expressions for the  deflection of light in both the strong field and weak field limits. Furthermore, we carried out a more detailed study of the images generated in the strong field limit, when the light passes very close to the photon sphere. In order to compare our results with those predicted by a Schwarzschild black hole, we modeled the solution with black hole data at the center of our galaxy and calculated the observables. We show that the angular separation increases with the LQG parameter, $a$, and, on the other hand, the relativistic images have increased intensity. We emphasize that this distinction between observables, generated by different models, is crucial for the investigation of gravitational theories beyond general relativity.
We further investigated the influence of the LQG parameter on the Einstein ring, concluding that for reasonable values of $a$, the radius and angular position of the ring differ significantly from the results predicted by the Schwarzschild solution. We hope that the present work, due to its analytical and numerical results, will contribute, in due time, to verifying the possibility of corrections to General Relativity.

\acknowledgments{The authors C.F.S. Pereira and R. L. L. Vitória would like to thank CAPES (Coordenação de Aperfeiçoamento de Pessoal de Nível Superior).}


\begin{thebibliography}{99}
	
\bibitem{1457}	Gravitational collapse and space-time singularities, Phys. Rev. Lett. {\bf 14}, 57 (1965).
	%
	
\bibitem{529}	S. W. Hawking, and R. Penrose, The singularities of gravitational collapse and cosmology, Proc. R Soc. Lond. Ser A. {\bf314},	529-548 (1970).
	
		\bibitem{tst} E. Berti et al., Class. Quant Grav. 32, 243001 (2015). 
		  \bibitem{mod-gra-rep}	T.~Clifton, P.~G.~Ferreira, A.~Padilla and C.~Skordis,
		  Phys.\ Rept.\  513, 1 (2012).
		  
		  \bibitem{fr theo}S. Capozziello and M. De Laurentis, Phys. Rept. 509, 167 (2011). 
		  
		  \bibitem{felice} A.~De Felice and S.~Tsujikawa,
		  Living Rev.\ Rel.\  13, 3 (2010).
		  
		  \bibitem{lb}F.~S.~N.~Lobo,
		  Dark Energy-Current Advances and Ideas
		  [arXiv:0807.1640 [gr-qc]].
		  
		   \bibitem{report} J.~Beltran Jimenez, L.~Heisenberg, G.~J.~Olmo and D.~Rubiera-Garcia,
		   Phys.\ Rept.\  {\bf 727}, 1 (2018).
		   
		   \bibitem{mod}S.~Nojiri, S.~D.~Odintsov and V.~K.~Oikonomou,
		   Phys.\ Rept.\  {\bf 692}, 1 (2017)
		   [arXiv:1705.11098 [gr-qc]].
	
	
		\bibitem{Rovelli003} C. Rovelli, Zakopane lectures on loop gravity, Proc. Sci., QGQGS2011 (2011) 003 [arXiv:1102.3660].
		
		\bibitem{Ashtekar84}   A. Ashtekar and E. Bianchi, A short review of loop quantum gravity, Rep. Prog. Phys. {\bf84}, 042001 (2021).
		
		\bibitem{Perez80}   A. Perez, Black holes in loop quantum gravity, Rep. Prog.
		Phys. {\bf80}, 126901 (2017).
		
		 \bibitem{Alonso-Bardaji829} A. Alonso-Bardají, Phys. Lett. B {\bf 829}, 137075 (2022).
		 
		 \bibitem{Alonso-Bardaji106} A. Alonso-Bardají, D. Brizuela, and R. Vera, Phys. Rev. D {\bf 106}, 024035 (2022).
		    
		
		      
		  \bibitem{Zhang08421} G. Fu, D. Zhang, P. Liu, X. M. Kuang and J. P. Wu, Peculiar properties in quasi-normal spectra from loop quantum gravity effect, arXiv:2301.08421. 
		      
		   \bibitem{Moreira107} Moreira, Zeus S, Junior, Haroldo C. D. Lima, Crispino, Lu\'{\i}s C. B, Herdeiro, Carlos A. R,  Phys. Rev. D {\bf107}, 104016 (2023).
		   
		  	\bibitem{Sobrinho14}  F. C. Sobrinho, H. A. Borges, I. P. R. Baranov, S. Carneiro, Classical Quantum Gravity {\bf40}, 145003
		   	(2023). 
		   		\bibitem{Einstein1936}  A. Einstein, Science {\bf84}, 506 (1936).
		   		
		   	\bibitem{Liebes} S. Liebes, Jr., Phys. Rev. {\bf133}, B835 (1964). 
		   		
		  	\bibitem{Darwin249} C. Darwin, Proc. R. Soc. A {\bf249}, 180 (1959); {\bf263}, 39 (1961).
		   \bibitem{Atkinson517}R. Atkinson, Astron. J. {\bf70}, 517 (1965).
		   	\bibitem{Virbhadra-Ellis-2000} K. S. Virbhadra and George F. R. Ellis,  Phys. Rev. D {\bf62}, 084003 (2000).
		   	
		   	
		   	
		   	\bibitem{Bozza2002} V. Bozza, Phys. Rev. D {\bf66}, 103001 (2002).
		   	
		   	\bibitem{Tsukamoto2017}  N. Tsukamoto, Phys.Rev. D {\bf95}, 064035 (2017).
		   	
		   	
		   		\bibitem{Eiroa2002} E. F. Eiroa, G. E. Romero, and D. F. Torres, Phys. Rev. D {\bf 66}, 024010 (2002). 
		   		
		   		\bibitem{Eiroa2004} E. F. Eiroa and D. F. Torres, Phys. Rev. D {\bf 69}, 063004 (2004).
		   		
		   		\bibitem{Tsukamoto201734} N.~Tsukamoto and Y.~Gong, Phys. Rev. D \textbf{95}, 064034 (2017). 
		   		
		   		
		   		
		   		
		   		\bibitem{rotacao} A. B. Aazami, C. R. Keeton, and A. O. Petters, J. Math. Phys. {\bf52}, 102501 (2011).
		   		
		   		\bibitem{Aazami} A. B. Aazami, C. R. Keeton, and A. O. Petters, J. Math. Phys. {\bf52}, 092502 (2011). 
		   		
		   		
		   		\bibitem{azquez}  S. V\'azquez and E.P. Esteban, Nuovo Cim. {\bf 119B}, 489 (2004).
		   		
		   		\bibitem{Bozza-Sereno}  V. Bozza, F. De Luca, G. Scarpetta, and M. Sereno,
		   		Phys. Rev. D {\bf 72}, 083003 (2005). 
		   		
		   		\bibitem{Bozza-Scarpetta} V. Bozza, F. De Luca, and G. Scarpetta, Phys. Rev. D {\bf 74}, 063001 (2006).
		   		
		   		\bibitem{VBz} V.Bozza, Phys. Rev. D {\bf67}, 103006 (2003). 
		   		
		   		\bibitem{Virbhadra-064038} Virbhadra, K. S, Phys. Rev. D {\bf 106}, 064038 (2022).
		   		
		   		\bibitem{Virbhadra-2204.01792}Virbhadra, K. S, Compactness of supermassive dark objects at galactic centers, 	arXiv:2204.01792.
		   		
		   		
		   		\bibitem{Nakajima85}  K. Nakajima and H. Asada, Phys. Rev. D {\bf 85}, 107501 (2012). 
		   		
		   		\bibitem{Chetouani} L. Chetouani and G. Cl{\'e}ment, Gen. Relativ. Gravit. {\bf 16}, 111 (1984).
		   		
		   		
		   		\bibitem{Nandi}  K. K. Nandi, Y. Z. Zhang, and A. V. Zakharov, Phys. Rev. D {\bf 74}, 024020 (2006). 
		   		
		   		\bibitem{Dey-sen}  T. K. Dey and S. Sen, Mod. Phys. Lett. A, {\bf 23}, 953 (2008). 
		   		
		   		
		   		\bibitem{Gibbons-Vyska}  G. W. Gibbons and M. Vyska, Class. Quant. Grav. {\bf 29}, 065016 (2012).
		   		
		   		
		   		\bibitem{Tsukamoto-Harada}   N. Tsukamoto, T. Harada, and K. Yajima, Phys. Rev. D {\bf 86}, 104062 (2012). 
		   		
		   		\bibitem{Nandi-Potapov}   K. K. Nandi, A. A. Potapov, R. N. Izmailov, A. Tamang, and J. C. Evans, Phys. Rev. D {\bf 93},	104044 (2016).
		   		
		   		\bibitem{Tsukamoto2017-95}   N. Tsukamoto, Phys. Rev. D {\bf95}, 084021 (2017). 
		   		
		   		\bibitem{Tsukamoto-Harada-2017}  N. Tsukamoto and T. Harada, Phys. Rev. D {\bf 95}, 024030 (2017).
		   		
		   		\bibitem{Shaikh-Banerjee}   R. Shaikh, P. Banerjee, S. Paul and T. Sarkar, JCAP {\bf1907}, 028 (2019). 
		   		
		   		\bibitem{Gao2211.17065} K. Gao, L.H. Liu, M. Zhu, Microlensing effects of wormholes associated to blackhole spacetimes. [arXiv:2211.17065 [grqc]]
		   		
		   		\bibitem{Huang2302.13704} T. Cai, Z. Wang, H. Huang, and M. Zhu, Higher order correction to weak-field lensing of Ellis-Bronnikov wormhole, arXiv:2302.13704 [gr-qc].
		   		
		   		\bibitem{Abe2010} F. Abe, Astrophys. J. 725, 787 (2010).
		   		
		   		
		   		\bibitem{Cheng} H. Cheng and J. Man,  Classical Quantum Gravity {\bf28}, 015001 (2011). 
		   		
		   		\bibitem{Sharif2015} M. Sharif, S. Iftikhar, Adv. High Energy Phys. {\bf2015}, 854264 (2015).
		   		
		   		\bibitem{Cheng92} J. Man, H. Cheng, Phys. Rev. D {\bf92}, 024004 (2015). 
		   		
		   		\bibitem{Cheng28} H. Cheng and J. Man, Class. Quant. Grav. {\bf28},015001 (2011). 
		   		
		   		\bibitem{Sotani92} H. Sotani and U. Miyamoto, Phys. Rev. D {\bf92}, 044052 (2015). 
		   		
		   		\bibitem{Wei75}  S. W. Wei, K. Yang, and Y. X. Liu, Eur. Phys. J. C {\bf 75}, 253 (2015) [Erratum: Eur. Phys. J. C {\bf} (2015) 331]. 
		   		
		   		\bibitem{Bhadra} A. Bhadra, Phys. Rev. D {\bf 67}, 103009 (2003).
		   		
		   		\bibitem{Eiroa73}  E.F . Eiroa, Phys. Rev. D {\bf73}, 043002 (2006).
		   		
		   		\bibitem{Sarkar23}   K. Sarkar, and A. Bhadra, Class. Quantum Grav. {\bf23}, 6101 (2006). 
		   		
		   		\bibitem{Mukherjee39}  N. Mukherjee, and A. S. Majumdar, Gen. Relativ. Gravit. {\bf39}, 583 (2007).
		   		
		   		\bibitem{Gyulchev75}  G. N. Gyulchev, and S. S. Yazadjiev, Phys. Rev. D {\bf75}, 023006 (2007). 
		   		
		   		\bibitem{Chen80}  S. Chen and J. Jing, Phys. Rev. D {\bf80}, 024036 (2009). 
		   		
		   		\bibitem{Shaikh96}  R. Shaikh and S. Kar, Phys. Rev. D {\bf96}, 044037 (2017). 
		   		
		   		
		   		\bibitem{Eiroa28} E. F. Eiroa and C. M. Sendra, Class. Quantum Grav. {\bf28}, 085008 (2011). 
		   		
		   		\bibitem{Eiroa88} E. F. Eiroa and C. M. Sendra, Phys. Rev. D {\bf88}, 103007 (2013).
		   		
		   		\bibitem{soaresBbounce} J. R. Nascimento, A. Y. Petrov, P. J. Porfirio, and  A. R. Soares, Phys. Rev. D {\bf102}, 044021 (2020).
		   		
		   		\bibitem {Ghosh006} S. Ghosh and A. Bhattacharyya, J. Cosmol. Astropart. Phys. {\bf 11}, 006, (2022).
		   		
		   	
		   	
		   	
		   	 \bibitem{AkiyamaL1} K. Akiyama et al., Astrophys. J. {\bf 875}, L1 (2019).
		   	 \bibitem{AkiyamaL2} K. Akiyama et al., Astrophys. J. Lett. {\bf 875}, L2 (2019).
		   	 \bibitem{AkiyamaL3} K. Akiyama et al., Astrophys. J. Lett. {\bf 875}, L3 (2019).
		   	 \bibitem{AkiyamaL4} K. Akiyama et al., Astrophys. J. Lett. {\bf 875}, L4 (2019).
		   	 \bibitem{AkiyamaL5} K. Akiyama et al., Astrophys. J. Lett. {\bf 875}, L5 (2019).
		   	 \bibitem{AkiyamaL6} K. Akiyama et al., Astrophys. J. Lett. {\bf 875}, L6 (2019).
		   	\bibitem{EHT2} https://www.ngeht.org/
		   	
		   	\bibitem{Boz-Cap2001} V. Bozza, S. Capozziello, G. Iovane, and G. Scarpetta, Gen. Rel. Grav. {\bf33}, 1535 (2001).
		   		\bibitem{Gezel2010} R. Genzel, F. Eisenhauer, S. Gillessen, Rev. Mod. Phys. {\bf 82}, 3121 (2010).
		   		 \bibitem{Bozza103005} V. Bozza, Phys. Rev. D {\bf78}, 103005 (2008). 
		   		 
		   		 \bibitem{Takizawa064060} K. Takizawa,T. Ono, H. Asada,
		   		 Phys. Rev. D {\bf 102}, 064060 (2020). 
\end{thebibliography}
\end{document}